\documentclass{PoS}

\title{A decomposition for SU(2) Yang-Mills fields}

\ShortTitle{A decomposition for SU(2) Yang-Mills fields}

\author{\speaker{Sedigheh Deldar}\\
        University of Tehran\\
        E-mail: \email{sdeldar@ut.ac.ir}}

\author{Ahmad Mohamadnejad\\
        University of Tehran\\
        E-mail: \email{a.mohamadnejad@ut.ac.ir}}
   
\abstract{Motivated by Abelian dominance, we suppose that the field strength tensor in the low energy limit of the SU(2) Yang-Mills theory is
$ \textbf{G}_{\mu\nu}=G_{\mu\nu} \textbf{n} $, where $ G_{\mu\nu} $ is a space-time tensor and $ \textbf{n} $ is a unit vector field
which selects the Abelian direction at each space-time point.
Based on this form of the field strength tensor, we propose a decomposition for the Yang-Mills field with three degrees of freedom.
It seems that by this kind of decompostion, both monopoles and vortices appear at the same time.
We have also obtained the Dirac quantization condition with a rescaled electric charge.}

\FullConference{Xth Quark Confinement and the Hadron Spectrum\\
                 8-12 October 2012\\
                 TUM Campus Garching, Munich, Germany}

\begin{document}
\section{Decomposition of the Yang-Mills field}
Even though the gauge field $ \textbf{A}_{\mu} $ is a proper order parameter for describing the Yang-Mills theory in its ultraviolet limit, the low energy limit of the Yang-Mills theory behaves like a dual superconductor \cite{Nambu} and
 some other order parameters may become more appropriate. Therefore, one can decompose the Yang-Mills fields to new collective variables. Decomposing the Yang-Mills fields has been done before by different methods \cite{Cho}. 
These decompositions pursue different purposes in particular in connection with the issue of quark confinement in quantum chromodynamics (QCD). 
We propose a decomposition based on the especial form of $ \textbf{G}_{\mu\nu} $ which is appropriate for the infrared regime of the SU(2) Yang-Mills theory with Abelian dominance \cite{Yotsuyanagi}
\begin{equation}
\textbf{G}_{\mu\nu}=G_{\mu\nu} \textbf{n}, \label{4}
\end{equation}
where $ G_{\mu\nu} $ is a colorless tensor and $ \textbf{n} $ is an isotriplet unit vector field which gives the Abelian direction at each
space-time point. One can construct an orthogonal basis for the color space by $ \textbf{n} $ and its derivatives, and then expand the gauge field $ \textbf{A}_{\mu} $ 
\begin{equation}
\textbf{A}_{\mu}=C_{\mu} \textbf{n} + \phi_{1} \partial_{\mu} \textbf{n} + \phi_{2}  \textbf{n}\times\partial_{\mu} \textbf{n}. \label{2}
\end{equation}
For SU(2) field strength tensor, we have
\begin{equation}
\textbf{G}_{\mu\nu} = \partial_{\mu} \textbf{A}_{\nu} - \partial_{\nu} \textbf{A}_{\mu} + g \textbf{A}_{\mu} \times \textbf{A}_{\nu}. \label{3}
\end{equation}
Substituting Eq. (\ref{2}) in (\ref{3}) and changing the variables $ \phi_{1} = \frac{\rho}{g^{2}} $, $ 1 + g \phi_{2} = \frac{\sigma}{g} $,
we get the following decomposition
\begin{equation}
\textbf{A}_{\mu}=C_{\mu} \textbf{n} + \frac{1}{g} \partial_{\mu} \textbf{n} \times \textbf{n}
+ \frac{\rho}{g^{2}} \partial_{\mu} \textbf{n} + \frac{\sigma}{g^{2}} \textbf{n}\times\partial_{\mu} \textbf{n}. \label{8}
\end{equation}
It is easy to show that Eq. (\ref{4}) is satisfied if $ C_{\mu} $ is decomposed as the following
\begin{eqnarray}
C_{\mu} = \frac{1}{g a^{2}} (\sigma \partial_{\mu} \rho - \rho \partial_{\mu} \sigma), \label{13}
\end{eqnarray}
A constraint on $ \rho $ and $ \sigma $ is obtained as well: $ \rho^{2} + \sigma^{2} = a^{2} $, $a$ is constant.
Therefore, $ C_{\mu} $ is restricted and it has one dynamical degree of freedom because it depends on
$ \rho $ and $ \sigma $ which are not independent. Thus, there are totally three dynamical degrees of freedom:
two for $ \textbf{n} $ and one for $ C_{\mu} $.

The field strength tensor can be rewritten in terms of electric and magnetic field strength tensors, $F_{\mu\nu}$ and $H_{\mu\nu}$, respectively:
\begin{equation}
\textbf{G}_{\mu\nu} = (F_{\mu\nu} + H_{\mu\nu}) \textbf{n}, \label{14}
\end{equation}
where
\begin{equation}
F_{\mu\nu} = \partial_{\mu} C_{\nu} - \partial_{\nu} C_{\mu}, \, \, \, 
H_{\mu\nu} = -\frac{1}{g^{\prime}} \textbf{n} .  (\partial_{\mu} \textbf{n} \times \partial_{\nu} \textbf{n}),\label{15}
\end{equation}
and $ \frac{1}{g^{\prime}} =  \frac{1}{g} - \frac{a^{2}}{g^{3}} $ where $ g^{\prime} $ is the rescaled electric charge.

\section{Vortices and monopoles}
Vortices will appear in  as topological singularities
of the scalar field $ (\rho,\sigma) $, if  "two-dimensional hedgehog ansatz" is chosen
\begin{equation}
(\rho,\sigma)=a\frac{\overrightarrow{r}}{r}=a (cos(m\varphi),sin(m\varphi)), \, \, \, \, \, \, m \in Z, \label{18}
\end{equation}
where $ \varphi $ is the azimuthal circular coordinate of $ S^{1}_{R} $.
Using Eq. (\ref{18}) in Eq. (\ref{13}) one obtains
\begin{eqnarray}
C_{\mu} = - \frac{m}{g} \partial_{\mu} \varphi
\Rightarrow C_{r} = C_{z} = 0 \, , \, \,\,\,\, C_{\varphi} = - \frac{m}{gr}  \Rightarrow
\overrightarrow{B} = -2m \frac{\delta(r)}{gr} \widehat{k}. \label{19}
\end{eqnarray}
This represents a vortex-like object which shows that  the magnetic field is singular on the z axis.

To obtain the magnetic monopole, we choose a hedgehog configuration $ \textbf{n} =\frac{r^{a}}{r} $
and we get,
\begin{eqnarray}
\overrightarrow{B} = B \widehat{r}, 
\, \, \, B = H_{\theta\varphi} = - \frac{1}{g^{\prime}} \textbf{n} . (\partial_{\theta} \textbf{n} \times \partial_{\varphi} \textbf{n}) 
 =  - \frac{m}{g^{\prime}} \frac{1}{r^{2}}, \, \, \, \, \, \, m \in Z \, . \label{31}
\end{eqnarray}
From Eq. (\ref{31}) one gets $ g^{\prime}  g_{m} = m $.
This is the Dirac quantization condition, but with a rescaled electric charge $ g^{\prime} $. 
For our decomposition, the relation between $ g $ and $ g_{m} $ is
\begin{equation}
g  g_{m} = m ( 1 - (\frac{a}{g})^{2} ). \label{35}
\end{equation}
If $ a $ goes to zero, the familiar Dirac quantization condition would be restored.
In addition, in this limit our decomposition reduces to the Cho decomposition where the U(1) field is no longer decomposed, and consequently no vortices appear.
So, it seems that the presence of vortices influences the Dirac quantization condition by rescaling the electric coupling or charge.

\section{Conclusion}
We conjecture a special form of the field strength tensor for the infrared limit of the SU(2) Yang-Mills theory to propose a decomposition for the Yang-Mills field. In this decomposition both vortices and monopoles can appear at the same time. Dirac quantization condition is also obtained, but with a rescaled color charge.

\end{document}